\documentclass[pre,aps,showpacs,twocolumn,superscriptaddress]{revtex4}
\usepackage{graphicx}
\usepackage{amsmath,amssymb}
\usepackage[utf8]{inputenc}
\begin{document}
\title{Hydrodynamic synchronization of autonomously oscillating optically
  trapped particles}
\author{Ivna Kavre}
\email{ivnakavre@gmail.com}
\affiliation{Faculty of Mathematics and Physics, University of Ljubljana,
  Jadranska 19, 1000 Ljubljana, Slovenia} 
\author{Andrej Vilfan}
\email{andrej.vilfan@ijs.si}
\affiliation{J. Stefan Institute, Jamova 39, 1000 Ljubljana, Slovenia}
\author{Dušan Babič}\email{dusan.babic@fmf.uni-lj.si}
\affiliation{Faculty of Mathematics and Physics, University of Ljubljana,
  Jadranska 19, 1000 Ljubljana, Slovenia} 
\date{\today}

\begin{abstract}
  Ellipsoidal micron-sized colloidal particles can oscillate spontaneously
  when trapped in a focused laser beam. If two oscillating particles are held
  in proximity their oscillations synchronize through hydrodynamic
  interactions. The degree of synchronization depends on the distance between
  the oscillators and on their orientation.  Due to the anisotropic nature of
  hydrodynamic coupling the synchronization is strongest when particles are
  arranged along the direction of oscillations. Similar behavior is observed
  for many oscillating particles arranged in a row.  Experimental observations
  are well reproduced with a model that uses a phenomenological description of
  the optical force and hydrodynamic interactions. Our results show that
  oscillating ellipsoidal particles can serve as a model system for studying
  hydrodynamic synchronization between biological cilia.
\end{abstract}


\maketitle 

Synchronization appears at all length scales from atoms to macroscopic bodies
and is ubiquitous in living systems \cite{Pikovsky2001}. For example at
microscale it plays an important role in the motion of microswimmers such as
protozoa \cite{Machemer1972}, algae \cite{Polin.Goldstein2009} or spermatozoa
\cite{Woolley.Revell2009} which use flagella or cilia for their propulsion. A
green alga \textit{Chlamydomonas} swims with two flagella which it moves in a
coordinated way reminiscent of a human doing the breaststroke. The
coordination of flagella is crucial for swimming along a straight line
\cite{Polin.Goldstein2009}.  Ciliates like \textit{Paramecium} coordinate the
beating of their cilia to form metachronal waves which greatly enhance their
swimming efficiency \cite{Osterman.Vilfan2011}. Similar waves are also
observed in cilia that cover respiratory epithelia and clear mucus from the
airways \cite{Gheber.Priel1994}. Hydrodynamic synchronization can even be
observed between flagella of two separate sperm cells swimming beside each
other \cite{Woolley.Revell2009}. In large numbers hydrodynamically
synchronized spermatozoa can form intriguing vortex patterns
\cite{Riedel.Howard2005}. The formation of metachronal waves has been
recreated in an artificial system consisting of microtubule bundles and
kinesin motors \cite{Sanchez.Dogic2011}.

Synchronization of nearby swimming microorganisms with waving tails was
first studied by G. Taylor who showed that it can lead to a reduced
dissipation \cite{Taylor1951}. However, this finding does not explain the
kinematic mechanism that keeps the tails synchronized. A difficulty lies in
the temporal reversibility of the Stokesian hydrodynamics while
synchronization is irreversible by nature \cite{Golestanian.Uchida2011}. This
can be overcome either by breaking the temporal symmetry in the driving
mechanism \cite{vilfan2006a,Uchida.Golestanian2011} or by describing a cilium
with more than one degree of freedom \cite{Niedermayer.Lenz2008}. In addition,
synchronization can also result from indirect coupling between cilia caused by
a rocking motion of the cell body \cite{Friedrich.Julicher2012,Geyer.Friedrich2013}, or from inertial effects at
non-zero Reynolds number \cite{Theers.Winkler2013}.

Hydrodynamic synchronization was studied in model system containing spherical
colloidal particles actively driven along closely spaced circular trajectories
via feedback controlled optical tweezers
\cite{Kotar.Cicuta2010,Damet.Cicuta2012,Bruot.Cicuta2012,Cicuta.Cicuta2012,Kotar.Cicuta2013}.
Cicuta and co-workers demonstrated an in-phase and anti-phase synchronization
where the transition between the two modes was controlled by changing the
shape of the driving potential \cite{Bruot.Cicuta2012}. In all these
experiments the driving mechanism required video particle tracking and
computer controlled feedback.

Alternatively hydrodynamic synchronization can be studied in a system of
optically trapped non-spherical particles. A recent study reported synchronous
rotation of two particles in vortex beams \cite{Arzola.Zemanek2014}. Elongated
particles in a focused beam can also exert oscillatory motion
\cite{Mihiretie.Pouligny2012,Mihiretie.Pouligny2013}. Their oscillations are
induced by a combination of the gradient force that pulls the particle towards
the beam center and a non-conservative force due to the radiation pressure.
In this Letter we report on experiments in which such oscillating ellipsoidal
particles were exploited to study the hydrodynamic synchronization between two
or more closely spaced autonomous oscillators.

\begin{figure}
  \begin{center}
    \includegraphics[width=8.6cm]{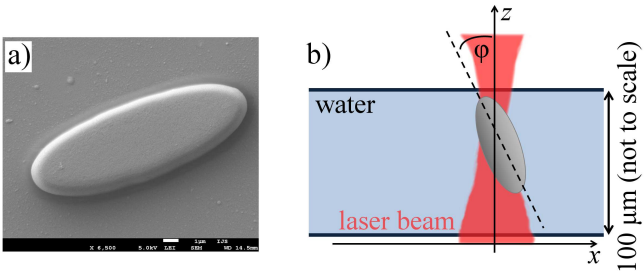}
  \end{center}
  \caption{a) SEM image of a particle with major radius $8\,\mu
    \rm m$, minor radius $2.5\, \mu \rm m$ and thickness $2.2\,\mu \rm m$. b)
    $\varphi$ denotes the angle between the particle's long axis and the
    $z$-axis, $x$ the distance between its center of reaction and the center
    of the laser beam.\label{fig:1}}
\end{figure}

\textit{Experiment.} Ellipsoidal prism shaped particles (Fig.~\ref{fig:1}a)
were fabricated from a layer of photoresist (SU-8, MicroChemicals) by maskless
photo-lithography using direct laser structuring (LPKF ProtoLaserD)
\cite{Kavcic.Poberaj2012}. Particles were dispersed in water and sealed in a
thick sample cell. The experiment was conducted on a setup built around an
inverted microscope (Zeiss Axiovert) equipped with a multi-trap laser tweezers
system (Aresis Tweez) using 1064 nm IR laser (Coherent Compass) and high
numerical aperture water immersion objective (Zeiss Achroplan $63\times
0.9\,\rm NA$) \cite{Kotar.Poberaj2006}.  Kinematic information on particle
motion was obtained from video recordings captured with a CMOS camera
(Pixelink PL-B 741) and subsequently analyzed by proprietary particle tracking
software \cite{Osterman2010}.  At the beginning of each measurement one or
multiple particles sedimented at the bottom of the experimental cell were
trapped and pushed against the upper cell wall. In order to control
  the orientation of a particle each trap consisted of a principal beam and a
  secondary beam with $20\%$ of the main beam's power that was shifted against
  the main beam by $2\,\rm \mu m$ perpendicular to the $x$-axis.

\begin{figure}
  \begin{center}
    \includegraphics[width=8.6cm]{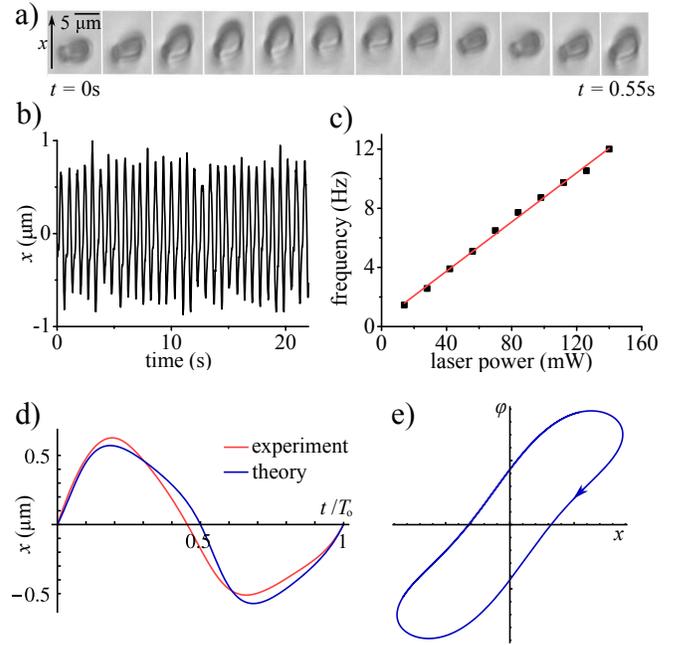}
  \end{center}
  \caption{a) Time sequence of an oscillatory motion of an
    ellipsoidal particle trapped by laser beam. b) Dynamics of the oscillating
    particle in y direction. c) Particle’s oscillation frequency as a function
    of laser power. The solid line represents a linear fit to the data. d)
    Oscillatory period of an ellipsoidal particle comparing the experimental
    and theoretical data for $\alpha=0.8$ and $\beta=1.5$. e) Stable limit
    cycle for same values of $\alpha$ and $\beta$ as in d).\label{fig:2}}
\end{figure}

\textit{Single particle oscillations.} Once a trapped particle is pushed
against the upper cell wall (Fig.~\ref{fig:1}b) it reaches a steady state in
which it persistently oscillates about the center of the trapping laser
beam. The oscillations take place in a plane orthogonal to the flat
  surface of the particle. Figure \ref{fig:2}a shows a time series of images
demonstrating the oscillatory motion. The position of the particle center as a
function of time is shown in Fig.~\ref{fig:2}b. Since all optical forces are
proportional to the light intensity we expected and found a linear dependence
of the oscillation frequency on the laser power (Fig.~\ref{fig:2}c).

\begin{figure}
  \begin{center}
    \includegraphics[width=6cm]{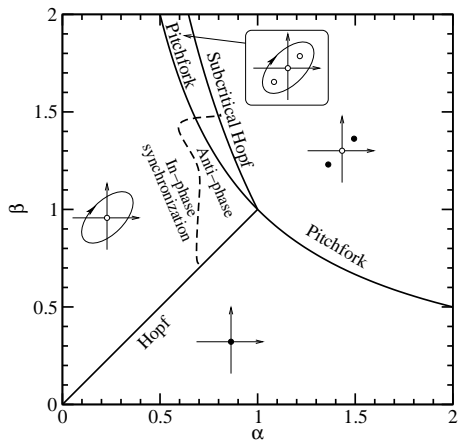}
  \end{center}
  \caption{Bifurcation diagram for a single particle described by
    Eqs.~\ref{eq:3}. Depending on the parameters $\alpha$ and $\beta$, the
    model can exhibit one or three fixed points and a limit cycle that
    describes autonomous oscillations. In addition, the diagram shows (dashed
    line) the regions in which two particles show in-phase and anti-phase
    synchronization for parameter values corresponding to longitudinal
    arrangement at large distances ($c_{RR}/c_{TT}=0.5$, $c_{TR}/(\epsilon
    c_{TT})=0.5$, $c_{RT}\epsilon/c_{TT}=1.5$).}
  \label{fig:3}
\end{figure}

The dynamics of an ellipsoidal particle in the optical tweezers beam can be
described as follows.  If we define the particle position as its center of
reaction \cite{Happel.Brenner} its mobility tensor becomes diagonal (a force
causes no rotation and a torque no translation) and the equations of motion
are
\begin{equation}
  \label{eq:2}
  \dot x= M_{TT} F\;, \qquad \dot \varphi=M_{RR} T 
\end{equation}
where $\varphi$ is the angle between the particle's long axis and the $z$-axis
and $x$ the distance between the center of reaction and the axis of the laser
beam (Fig.~\ref{fig:1}b). We propose the following expressions for the force
and the torque acting on the particle
\begin{equation}
  \label{eq:1}
\begin{split}
  F&=-A x + B \varphi \\
  T&= -C x +D \varphi -E \varphi^3\;.
\end{split}
\end{equation}
The coefficient $A$ describes the restoring potential of the laser trap, $B$
is the effect of tilt -- the radiation pressure on a tilted particle exerts a
lateral force; $C$ is the restoring torque of the trap boundary; $D$ describes
the torque on a particle pressed against a wall that causes an instability and
$E$ is a restoring torque that counteracts the instability.  The equations of
motion (\ref{eq:2}) can then be written in the form
\begin{equation}
  \label{eq:3}
  \begin{split}
    \dot x&=\Omega (-\alpha x +\epsilon \varphi)\\
    \dot \varphi & = \Omega( -\frac 1 \epsilon x +\beta \varphi -\delta \varphi^3)\;.
  \end{split}
\end{equation}
Of the five parameters $\Omega$ determines the oscillation frequency,
$\delta$ the amplitude and $\epsilon$ the amplitude ratio between $x$ and
$\varphi$. The dynamical behavior of the system is governed by the two essential
parameters ($\alpha$ and $\beta$).
The dynamical system described by Eq.~\ref{eq:3} always has a fixed point at
$(0,0)$ and for $\alpha\beta>1$ it has two more at $\pm \sqrt{(\alpha \beta
  -1)/(\alpha \delta)} (\epsilon/\alpha,1)$. The fixed point at $(0,0)$ is
stable if $\alpha>\beta$ and $\alpha \beta<1$ . At $\alpha=\beta$ it undergoes
a Hopf bifurcation and becomes unstable, encircled by a limit cycle. At
$\alpha\beta=1$ there is a pitchfork bifurcation (transition from 1 to 3 fixed
points) and for $\alpha\beta>1$ $(0,0)$ becomes a saddle point. The remaining
two fixed points are stable if $\alpha+2\beta-3/\alpha>0$ and unstable
otherwise. The latter transition is a subcritical Hopf bifurcation. The
dynamical regimes of the system are shown in Fig.~\ref{fig:3}.

Figure \ref{fig:2}d shows a measured oscillation cycle obtained by averaging
the experimentally obtained particle position as a function of time over many
periods. The shape of this oscillation cycle is visibly asymmetric and
compares well to the trajectories obtained from model equations in the region
with 3 fixed points.  Equations of motion (\ref{eq:3}) were solved numerically
for $\alpha=0.8$ and $\beta=1.5$. The limit cycle of the oscillator for the
same parameters is shown in Fig.~\ref{fig:2}e.  From this comparison we
conclude that an oscillating particle can be adequately described by the
phenomenological model we introduced here. Furthermore, it has been
demonstrated that an oscillator with two degrees of freedom is a good
candidate for hydrodynamic synchronization in \cite{Niedermayer.Lenz2008}.

\begin{figure}
  \begin{center}
    \includegraphics[width=8.6cm]{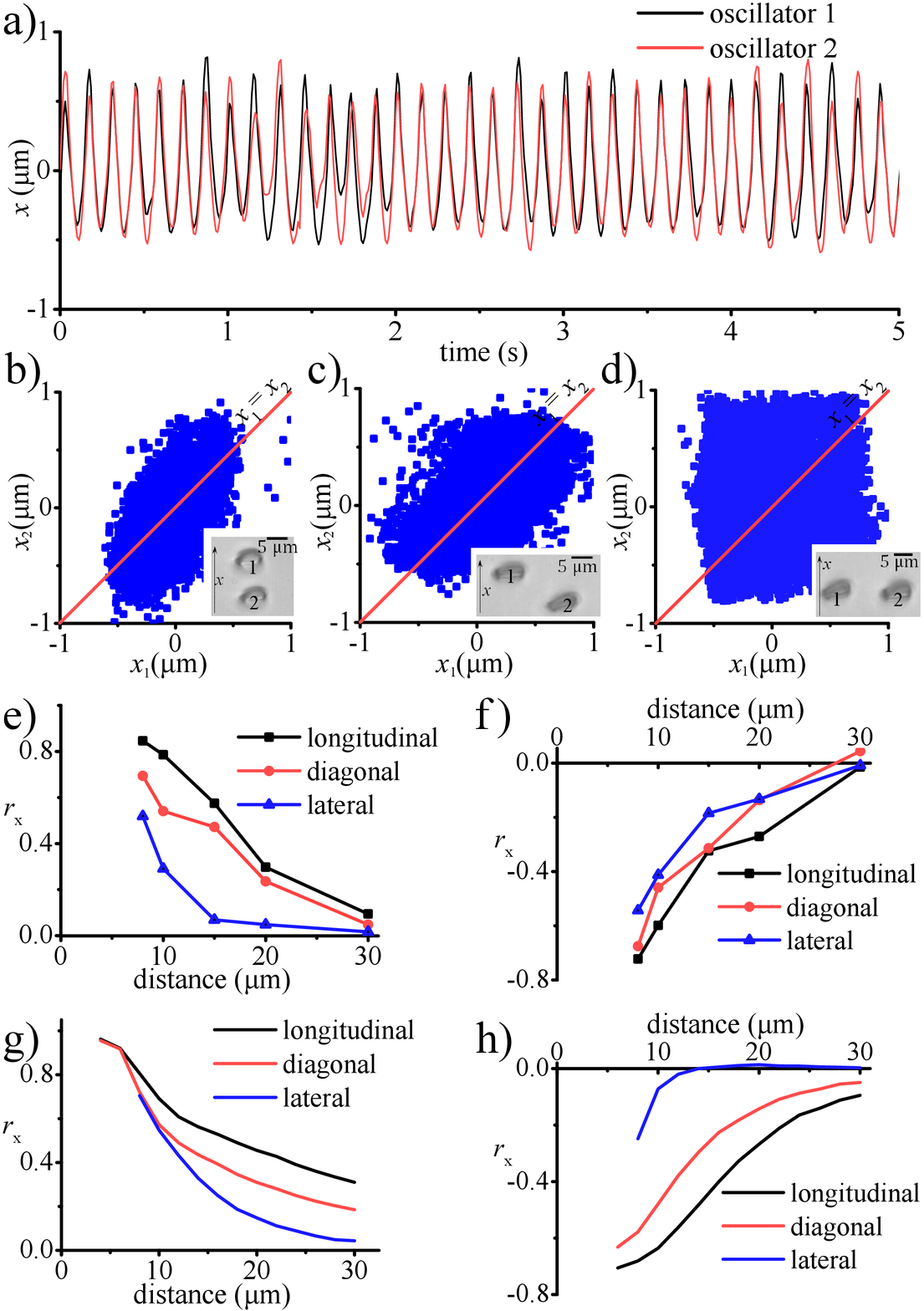}
  \end{center}
  \caption{a) Position oscillations of two particles in
    longitudinal arrangement (inset in b)) at a distance of $8\,\mu \rm m$,
    laser power 56 mW. b)-d) Synchronization strength at distance $15\,\mu \rm
    m$ for three arrangements: b) longitudinal, c) diagonal, d) lateral. The
    insets show the position of the particles. Laser power: 56 mW. e)-f)
    Correlation coefficient $r_x$ as a function of the interparticle distance
    $d$ for different arrangements showing (e) in-phase synchronization (laser
    power 28 mW) and (f) anti-phase synchronization (laser power 42 mW). g)-h)
    Theoretically calculated correlation coefficients with parameters
    $\alpha=0.8$, $\beta=1.5$, $\epsilon=14\,\rm \mu m$ (g) and
    $\epsilon=6\,\rm \mu m$ (h).\label{fig:4}}
\end{figure}

\textit{Two particle synchronization.} To study hydrodynamically mediated
synchronization between independent oscillators we trapped two ellipsoidal
particles in separate beams and brought them in close proximity. Measurements
were performed for three different alignments of the particles with respect to
the direction of oscillation: longitudinal (inset Fig.~\ref{fig:4}b), diagonal
(Fig.~\ref{fig:4}c) and lateral alignment (Fig.~\ref{fig:4}d). An example of particle
positions vs.\ time is shown in Fig.~\ref{fig:4}a for an interparticle
distance of $8 \,\mu \rm m$ (half the particle size) and longitudinal
configuration. Synchronized oscillation is clearly visible.

The degree of synchronization can be visualized in a scatter plot of particle
positions $x_1$ vs.\ $x_2$ (Fig.~\ref{fig:4}b-d).  For an interparticle
distance of $15 \,\mu \rm m$ the graph shows a strongly correlated motion for
longitudinal alignment (Fig.~\ref{fig:4}b) indicating that oscillating
particles synchronized in-phase.  For diagonal alignment the motion is still
synchronous although the correlation is weaker (Fig.~\ref{fig:4}c). For the
lateral alignment the motion is not correlated showing that the oscillations
are out of synchrony (Fig.~\ref{fig:4}d).  This observation is in accordance
with the anisotropy of Oseen's tensor which is even more pronounced in the
presence of a no-slip boundary \cite{Blake.1971}.  We quantitatively evaluate
the degree of synchrony by calculating the Pearson correlation coefficient as
a function of interparticle distance (Fig.~\ref{fig:4}e). In one instance the
particles also showed robust antiphase synchronization (Fig.~\ref{fig:4}f). The type of
synchronization was largely invariant of the laser power. We conclude that the
transition from in-phase to anti-phase synchronization was governed by the
detailed properties of the laser beam in the vicinity of the surface.

To understand the hydrodynamic synchronization we extend the phenomenological
model (\ref{eq:3}) by taking into account the fluid-mediated interaction. The
force or torque which sets one particle and the surrounding fluid in motion
also has an effect on the velocity and angular velocity of the second
particle. Their equations of motion can be written as
\begin{equation}
  \label{eq:interactions}
  \begin{split}
    \dot x_{1,2}=M_{TT} F_{1,2}+ C_{TT} F_{2,1} + C_{TR} T_{2,1}\\
    \dot \varphi_{1,2}=M_{RR} T_{1,2}+ C_{RT} F_{2,1} + C_{RR} T_{2,1}
  \end{split}
\end{equation}
where $x_i$ and $\varphi_i$ denote the deflection of $i$-th ($i=1,2$)
particle, $F_i$ the force and $T_i$ the torque acting on it. $C_{TT}$,
$C_{TR}$, \ldots\ denote the coupling coefficients 
(off-diagonal elements of the two-particle grand mobility matrix). By
introducing the reduced coefficients $c_{TT}=C_{TT}/M_{TT}$,
$c_{RR}=C_{RR}/M_{RR}$, $c_{TR}= C_{TR}/M_{RR}$, $c_{RT}=C_{RT}/M_{TT}$ and
inserting $F$ and $T$ from Eq.~\ref{eq:1} we obtain
\begin{equation}
  \label{eq:coupledeqs}
  \begin{split}
    \dot x_{1}&=\Omega_{1}(-\alpha x_{1} +\epsilon \varphi_{1}) +c_{TT}
    \Omega_{2}(-\alpha x_{2} +\epsilon \varphi_{2}) \\&+c_{TR} \Omega_{2}(
    -\frac 1 \epsilon x_{2} +\beta \varphi_{2} -\delta \varphi_{2}^3)\\
    \dot \varphi_{1} & = \Omega_{1}( -\frac 1 \epsilon x_{1} +\beta \varphi_{1} -\delta
    \varphi_{1}^3)+c_{RT}
    \Omega_{2}(-\alpha x_{2} +\epsilon \varphi_{2}) \\&+c_{RR} \Omega_{2}(
    -\frac 1 \epsilon x_{2} +\beta \varphi_{2} -\delta \varphi_{2}^3)
  \end{split}
\end{equation}
for particle $1$ and analogous equations for particle $2$.  In this expression
we assumed that the two oscillators differ only in their characteristic
frequencies ($\Omega_1$ and $\Omega_2$).  A numerical solution of the
equations of motion (\ref{eq:coupledeqs}) for two identical oscillators
($\Omega_1=\Omega_2$) reveals that if the particles are only coupled through
the $c_{TT}$ coefficient, they will synchronize in anti-phase except
for a narrow region with small $\alpha$. The coefficients $c_{RR}$ and
$c_{RT}$ always lead to in-phase synchronization and $c_{TR}$ always to
anti-phase. As a consequence, the nature of synchronization will depend
sensitively on the ratio between coupling coefficients.

We determined the coupling coefficients numerically by modeling the particles
as ellipsoids orthogonal to a wall. We solved the hydrodynamic problem with a
boundary element method (BEM) by representing each particle with 512 elements
and incorporating the no-slip boundary condition at the plate into the Green
function \cite{Pozrikidis2002} (Fig.~\ref{fig:s1}). We assumed a distance between the particle tip and the wall of $0.1
\,\mu \rm m$ and checked that the choice of this value only has a minor
influence on the calculated coefficients. In the calculation we
  restricted the motion to the direction along the $x$-axis in which active
  oscillations take place. The resulting coupling coefficients are shown in
Fig.~\ref{fig:s2}.  In order to reproduce a
realistic situation we simulated two oscillators with a frequency
  mismatch $\Omega_2/\Omega_1=1.2$ (based on the measured average frequency
  difference between particles that are out of range of hydrodynamics
  interactions) and additional noise. The resulting correlation coefficients
are shown in Fig.~\ref{fig:4}g. They show qualitative agreement with the
measured values under similar conditions, both in their distance- and
direction-dependence.  Frequency mismatch and noise both contribute to
  the loss of synchrony at larger distances. The relative weight of different
coupling coefficients $c_{TT}$ etc.\ depends sensitively on the parameter
$\epsilon$. A small variation can lead to a regime where the two oscillators
show anti-phase synchronization (Fig.~\ref{fig:4}h). As one can see from the
phase diagram for typical parameters (Fig.~\ref{fig:3}, dashed line) the
parameter range that allows anti-phase synchronization is rather narrow. This
provides a possible explanation for the rare instances in which anti-phase
synchronization was observed in the experiment.

\textit{Synchronization in particle row}. Following the two-particle
experiments we investigated if hydrodynamic interactions can also lead to
synchronization in a row of particles.  When biological cilia are densely
covering a surface their oscillations usually form metachronal waves whose
direction and wavelength vary from system to system. A problem that arises in
theoretical models of wave formation is that the cilia at the ends of a chain
are subject to different interactions with their neighbors than those in the
middle. As a consequence some models obtain synchronization or metachronal
waves when they introduce periodic boundary conditions but not in a finite
chain \cite{Lenz.Ryskin2006}, although the latter is possible, too
\cite{Wollin.Stark2011}. Likewise, experimental studies on model systems
concentrated on particles arranged in closed ring structures
\cite{Damet.Cicuta2012}.

\begin{figure}
  \begin{center}
    \includegraphics[width=8.6cm]{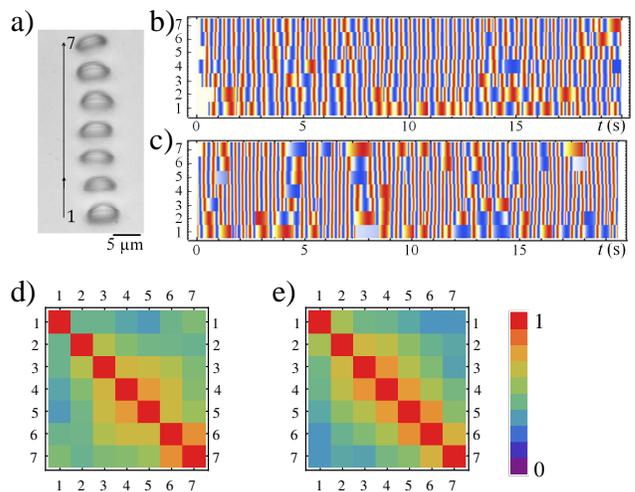}
  \end{center}
  \caption{a) Particles arranged in a row. b) Phase profile:
    laser power 40 mW per particle and distance between adjacent particles
    $8$-$9\,\mu \rm m$. c) Simulation of the same system using parameters
    $\alpha=0.8$, $\beta=1.5$, $\epsilon=12\,\rm \mu m$. d) Correlation matrix
    of experimentally measured particle positions. e) Equivalent correlation
    matrix from simulation. \label{fig:row}}
\end{figure}

In our experiment 7 optical traps were arranged in a linear row with a spacing
of $8\,\mu \rm m$ (Fig.~\ref{fig:row}a).  Particles never showed complete synchrony or
phase waves. Figure \ref{fig:row}b shows an example where a high degree of
local synchrony persisted over many oscillations but not indefinitely. The
correlations are stronger in the middle of the chain and weaker at its ends
(Fig.~\ref{fig:row}d).

We also extended the theoretical model to a row of particles. We used the same
parameters as for two particles and applied the distance-dependent coupling
coefficients (Fig.\ S2) to describe the interaction between a particle and all
other particles in the chain.  A phase plot of a simulated system with the
same parameters we used for two particles is shown in Fig.~\ref{fig:row}c and
the correlation matrix between all oscillators in Fig.~\ref{fig:row}e. Like in
the experiment partial order but no complete synchronization can be seen. The
loss of synchrony is mainly due to geometric effects in the chain, because the
particles at the end are subject to weaker interactions than those in the
middle.

In conclusion, we have shown that elliptical colloidal particles oscillating in
a laser beam provide a good model system for studying hydrodynamic
synchronization. In a two-particle system we showed that the coupling strength
depends on the interparticle distance as well as their spatial
arrangement. While particles usually synchronize in-phase, anti-phase
synchronization was observed as well.  For many particles arranged in a row we
observe only weak synchronization in agreement with theoretical results. A
question that remains to be investigated is whether 2-dimensional arrays of
oscillators would show more robust synchronization and what are the necessary
conditions for the formation of metachronal waves in such a system.

\acknowledgments 
We thank Dr.\ Peter Panjan for preparing the nickel coated substrates. This
work has been supported financially by the European Union Seventh Framework
Program (FP7) ITN Marie Curie and the Slovenian Research Agency, grant
J1-5437.

\renewcommand{\thefigure}{S\arabic{figure}}
\setcounter{figure}{0}
\begin{figure}
    \includegraphics[width=6cm]{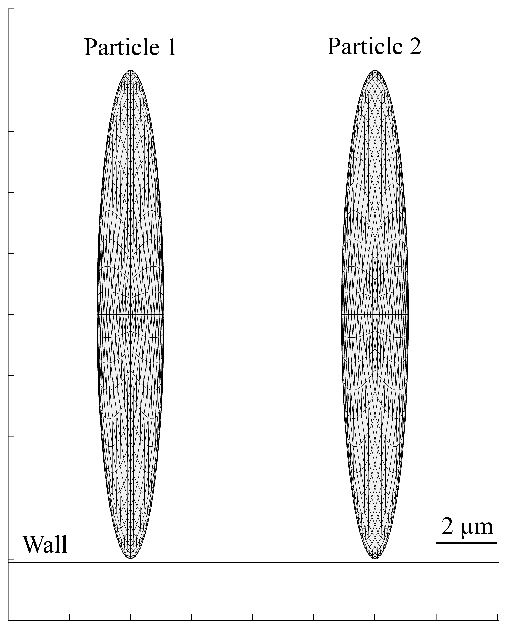}
\caption{Boundary element representation of two ellipsoidal particles (side
  view) in close proximity to a wall with no-slip boundary condition.}
\label{fig:s1}
\end{figure}
\begin{figure*}
  \begin{center}
    \includegraphics[width=14cm]{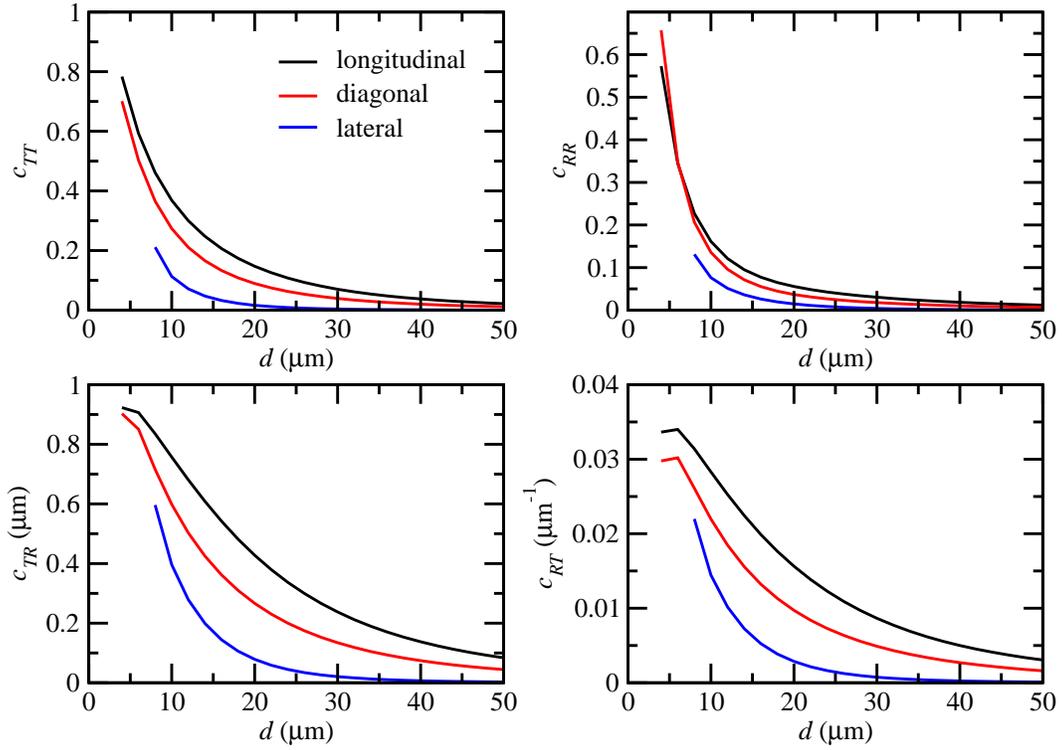}
  \end{center}
\caption{Reduced coupling coefficients $c_{TT}$, $c_{RR}$, $c_{TR}$, $c_{RT}$
  calculated with the boundary element method (BEM) as a function of
  interparticle distance $d$ for longitudinal, diagonal and lateral spatial
  arrangement.}
\label{fig:s2}
\end{figure*}


\begin{thebibliography}{33}
\expandafter\ifx\csname natexlab\endcsname\relax\def\natexlab#1{#1}\fi
\expandafter\ifx\csname bibnamefont\endcsname\relax
  \def\bibnamefont#1{#1}\fi
\expandafter\ifx\csname bibfnamefont\endcsname\relax
  \def\bibfnamefont#1{#1}\fi
\expandafter\ifx\csname citenamefont\endcsname\relax
  \def\citenamefont#1{#1}\fi
\expandafter\ifx\csname url\endcsname\relax
  \def\url#1{\texttt{#1}}\fi
\expandafter\ifx\csname urlprefix\endcsname\relax\def\urlprefix{URL }\fi
\providecommand{\bibinfo}[2]{#2}
\providecommand{\eprint}[2][]{\url{#2}}

\bibitem[{\citenamefont{Pikovsky et~al.}(2001)\citenamefont{Pikovsky,
  Rosenblum, and Kurths}}]{Pikovsky2001}
\bibinfo{author}{\bibfnamefont{A.}~\bibnamefont{Pikovsky}},
  \bibinfo{author}{\bibfnamefont{M.}~\bibnamefont{Rosenblum}},
  \bibnamefont{and} \bibinfo{author}{\bibfnamefont{J.}~\bibnamefont{Kurths}},
  \emph{\bibinfo{title}{Synchronization, {A} Universal Concept in Nonlinear
  Sciences}} (\bibinfo{publisher}{Cambridge Univeristy Press},
  \bibinfo{address}{Cambridge}, \bibinfo{year}{2001}).

\bibitem[{\citenamefont{Machemer}(1972)}]{Machemer1972}
\bibinfo{author}{\bibfnamefont{H.}~\bibnamefont{Machemer}},
  \bibinfo{journal}{J. Exp.\ Biol.} \textbf{\bibinfo{volume}{57}},
  \bibinfo{pages}{239} (\bibinfo{year}{1972}).

\bibitem[{\citenamefont{Polin et~al.}(2009)\citenamefont{Polin, Tuval,
  Drescher, Gollub, and Goldstein}}]{Polin.Goldstein2009}
\bibinfo{author}{\bibfnamefont{M.}~\bibnamefont{Polin}},
  \bibinfo{author}{\bibfnamefont{I.}~\bibnamefont{Tuval}},
  \bibinfo{author}{\bibfnamefont{K.}~\bibnamefont{Drescher}},
  \bibinfo{author}{\bibfnamefont{J.~P.} \bibnamefont{Gollub}},
  \bibnamefont{and} \bibinfo{author}{\bibfnamefont{R.~E.}
  \bibnamefont{Goldstein}}, \bibinfo{journal}{Science}
  \textbf{\bibinfo{volume}{325}}, \bibinfo{pages}{487} (\bibinfo{year}{2009}).

\bibitem[{\citenamefont{Woolley et~al.}(2009)\citenamefont{Woolley, Crockett,
  Groom, and Revell}}]{Woolley.Revell2009}
\bibinfo{author}{\bibfnamefont{D.~M.} \bibnamefont{Woolley}},
  \bibinfo{author}{\bibfnamefont{R.~F.} \bibnamefont{Crockett}},
  \bibinfo{author}{\bibfnamefont{W.~D.} \bibnamefont{Groom}}, \bibnamefont{and}
  \bibinfo{author}{\bibfnamefont{S.~G.} \bibnamefont{Revell}},
  \bibinfo{journal}{J Exp Biol} \textbf{\bibinfo{volume}{212}},
  \bibinfo{pages}{2215} (\bibinfo{year}{2009}).

\bibitem[{\citenamefont{Osterman and Vilfan}(2011)}]{Osterman.Vilfan2011}
\bibinfo{author}{\bibfnamefont{N.}~\bibnamefont{Osterman}} \bibnamefont{and}
  \bibinfo{author}{\bibfnamefont{A.}~\bibnamefont{Vilfan}},
  \bibinfo{journal}{Proc.\ Natl.\ Acad.\ Sci.\ USA}
  \textbf{\bibinfo{volume}{108}}, \bibinfo{pages}{15727}
  (\bibinfo{year}{2011}).

\bibitem[{\citenamefont{Gheber and Priel}(1994)}]{Gheber.Priel1994}
\bibinfo{author}{\bibfnamefont{L.}~\bibnamefont{Gheber}} \bibnamefont{and}
  \bibinfo{author}{\bibfnamefont{Z.}~\bibnamefont{Priel}},
  \bibinfo{journal}{Cell~Motil.~Cytoskeleton} \textbf{\bibinfo{volume}{28}},
  \bibinfo{pages}{333} (\bibinfo{year}{1994}).

\bibitem[{\citenamefont{Riedel et~al.}(2005)\citenamefont{Riedel, Kruse, and
  Howard}}]{Riedel.Howard2005}
\bibinfo{author}{\bibfnamefont{I.~H.} \bibnamefont{Riedel}},
  \bibinfo{author}{\bibfnamefont{K.}~\bibnamefont{Kruse}}, \bibnamefont{and}
  \bibinfo{author}{\bibfnamefont{J.}~\bibnamefont{Howard}},
  \bibinfo{journal}{Science} \textbf{\bibinfo{volume}{309}},
  \bibinfo{pages}{300} (\bibinfo{year}{2005}).

\bibitem[{\citenamefont{Sanchez et~al.}(2011)\citenamefont{Sanchez, Welch,
  Nicastro, and Dogic}}]{Sanchez.Dogic2011}
\bibinfo{author}{\bibfnamefont{T.}~\bibnamefont{Sanchez}},
  \bibinfo{author}{\bibfnamefont{D.}~\bibnamefont{Welch}},
  \bibinfo{author}{\bibfnamefont{D.}~\bibnamefont{Nicastro}}, \bibnamefont{and}
  \bibinfo{author}{\bibfnamefont{Z.}~\bibnamefont{Dogic}},
  \bibinfo{journal}{Science} \textbf{\bibinfo{volume}{333}},
  \bibinfo{pages}{456} (\bibinfo{year}{2011}).

\bibitem[{\citenamefont{Taylor}(1951)}]{Taylor1951}
\bibinfo{author}{\bibfnamefont{G.}~\bibnamefont{Taylor}},
  \bibinfo{journal}{Proc.\ R. Soc.\ Lond.\ A} \textbf{\bibinfo{volume}{209}},
  \bibinfo{pages}{447} (\bibinfo{year}{1951}).

\bibitem[{\citenamefont{Golestanian et~al.}(2011)\citenamefont{Golestanian,
  Yeomans, and Uchida}}]{Golestanian.Uchida2011}
\bibinfo{author}{\bibfnamefont{R.}~\bibnamefont{Golestanian}},
  \bibinfo{author}{\bibfnamefont{J.~M.} \bibnamefont{Yeomans}},
  \bibnamefont{and} \bibinfo{author}{\bibfnamefont{N.}~\bibnamefont{Uchida}},
  \bibinfo{journal}{Soft Matter} \textbf{\bibinfo{volume}{7}},
  \bibinfo{pages}{3074} (\bibinfo{year}{2011}).

\bibitem[{\citenamefont{Vilfan and J{\" u}licher}(2006)}]{vilfan2006a}
\bibinfo{author}{\bibfnamefont{A.}~\bibnamefont{Vilfan}} \bibnamefont{and}
  \bibinfo{author}{\bibfnamefont{F.}~\bibnamefont{J{\" u}licher}},
  \bibinfo{journal}{Phys.~Rev.~Lett.} \textbf{\bibinfo{volume}{96}},
  \bibinfo{pages}{058102} (\bibinfo{year}{2006}).

\bibitem[{\citenamefont{Uchida and Golestanian}(2011)}]{Uchida.Golestanian2011}
\bibinfo{author}{\bibfnamefont{N.}~\bibnamefont{Uchida}} \bibnamefont{and}
  \bibinfo{author}{\bibfnamefont{R.}~\bibnamefont{Golestanian}},
  \bibinfo{journal}{Phys.~Rev.~Lett.} \textbf{\bibinfo{volume}{106}},
  \bibinfo{pages}{058104} (\bibinfo{year}{2011}).

\bibitem[{\citenamefont{Niedermayer et~al.}(2008)\citenamefont{Niedermayer,
  Eckhardt, and Lenz}}]{Niedermayer.Lenz2008}
\bibinfo{author}{\bibfnamefont{T.}~\bibnamefont{Niedermayer}},
  \bibinfo{author}{\bibfnamefont{B.}~\bibnamefont{Eckhardt}}, \bibnamefont{and}
  \bibinfo{author}{\bibfnamefont{P.}~\bibnamefont{Lenz}},
  \bibinfo{journal}{Chaos} \textbf{\bibinfo{volume}{18}},
  \bibinfo{pages}{037128} (\bibinfo{year}{2008}).

\bibitem[{\citenamefont{Friedrich and
  Jülicher}(2012)}]{Friedrich.Julicher2012}
\bibinfo{author}{\bibfnamefont{B.~M.} \bibnamefont{Friedrich}}
  \bibnamefont{and}
  \bibinfo{author}{\bibfnamefont{F.}~\bibnamefont{Jülicher}},
  \bibinfo{journal}{Phys.~Rev.~Lett.} \textbf{\bibinfo{volume}{109}},
  \bibinfo{pages}{138102} (\bibinfo{year}{2012}).

\bibitem[{\citenamefont{Geyer et~al.}(2013)\citenamefont{Geyer, Jülicher,
  Howard, and Friedrich}}]{Geyer.Friedrich2013}
\bibinfo{author}{\bibfnamefont{V.~F.} \bibnamefont{Geyer}},
  \bibinfo{author}{\bibfnamefont{F.}~\bibnamefont{Jülicher}},
  \bibinfo{author}{\bibfnamefont{J.}~\bibnamefont{Howard}}, \bibnamefont{and}
  \bibinfo{author}{\bibfnamefont{B.~M.} \bibnamefont{Friedrich}},
  \bibinfo{journal}{Proc.\ Natl.\ Acad.\ Sci.\ USA}
  \textbf{\bibinfo{volume}{110}}, \bibinfo{pages}{18058}
  (\bibinfo{year}{2013}).

\bibitem[{\citenamefont{Theers and Winkler}(2013)}]{Theers.Winkler2013}
\bibinfo{author}{\bibfnamefont{M.}~\bibnamefont{Theers}} \bibnamefont{and}
  \bibinfo{author}{\bibfnamefont{R.~G.} \bibnamefont{Winkler}},
  \bibinfo{journal}{Phys.~Rev.~E} \textbf{\bibinfo{volume}{88}},
  \bibinfo{pages}{023012} (\bibinfo{year}{2013}).

\bibitem[{\citenamefont{Kotar et~al.}(2010)\citenamefont{Kotar, Leoni,
  Bassetti, Lagomarsino, and Cicuta}}]{Kotar.Cicuta2010}
\bibinfo{author}{\bibfnamefont{J.}~\bibnamefont{Kotar}},
  \bibinfo{author}{\bibfnamefont{M.}~\bibnamefont{Leoni}},
  \bibinfo{author}{\bibfnamefont{B.}~\bibnamefont{Bassetti}},
  \bibinfo{author}{\bibfnamefont{M.~C.} \bibnamefont{Lagomarsino}},
  \bibnamefont{and} \bibinfo{author}{\bibfnamefont{P.}~\bibnamefont{Cicuta}},
  \bibinfo{journal}{Proc.\ Natl.\ Acad.\ Sci.\ USA}
  \textbf{\bibinfo{volume}{107}}, \bibinfo{pages}{7669} (\bibinfo{year}{2010}).

\bibitem[{\citenamefont{Damet et~al.}(2012)\citenamefont{Damet, Cicuta, Kotar,
  Lagomarsino, and Cicuta}}]{Damet.Cicuta2012}
\bibinfo{author}{\bibfnamefont{L.}~\bibnamefont{Damet}},
  \bibinfo{author}{\bibfnamefont{G.~M.} \bibnamefont{Cicuta}},
  \bibinfo{author}{\bibfnamefont{J.}~\bibnamefont{Kotar}},
  \bibinfo{author}{\bibfnamefont{M.~C.} \bibnamefont{Lagomarsino}},
  \bibnamefont{and} \bibinfo{author}{\bibfnamefont{P.}~\bibnamefont{Cicuta}},
  \bibinfo{journal}{Soft Matter} \textbf{\bibinfo{volume}{8}},
  \bibinfo{pages}{8672} (\bibinfo{year}{2012}).

\bibitem[{\citenamefont{Bruot et~al.}(2012)\citenamefont{Bruot, Kotar, {de
  Lillo}, {Cosentino Lagomarsino}, and Cicuta}}]{Bruot.Cicuta2012}
\bibinfo{author}{\bibfnamefont{N.}~\bibnamefont{Bruot}},
  \bibinfo{author}{\bibfnamefont{J.}~\bibnamefont{Kotar}},
  \bibinfo{author}{\bibfnamefont{F.}~\bibnamefont{{de Lillo}}},
  \bibinfo{author}{\bibfnamefont{M.}~\bibnamefont{{Cosentino Lagomarsino}}},
  \bibnamefont{and} \bibinfo{author}{\bibfnamefont{P.}~\bibnamefont{Cicuta}},
  \bibinfo{journal}{Phys.~Rev.~Lett.} \textbf{\bibinfo{volume}{109}},
  \bibinfo{pages}{164103} (\bibinfo{year}{2012}).

\bibitem[{\citenamefont{Cicuta et~al.}(2012)\citenamefont{Cicuta, Onofri,
  Lagomarsino, and Cicuta}}]{Cicuta.Cicuta2012}
\bibinfo{author}{\bibfnamefont{G.~M.} \bibnamefont{Cicuta}},
  \bibinfo{author}{\bibfnamefont{E.}~\bibnamefont{Onofri}},
  \bibinfo{author}{\bibfnamefont{M.~C.} \bibnamefont{Lagomarsino}},
  \bibnamefont{and} \bibinfo{author}{\bibfnamefont{P.}~\bibnamefont{Cicuta}},
  \bibinfo{journal}{Phys.~Rev.~E} \textbf{\bibinfo{volume}{85}},
  \bibinfo{pages}{016203} (\bibinfo{year}{2012}).

\bibitem[{\citenamefont{Kotar et~al.}(2013)\citenamefont{Kotar, Debono, Bruot,
  Box, Phillips, Simpson, Hanna, and Cicuta}}]{Kotar.Cicuta2013}
\bibinfo{author}{\bibfnamefont{J.}~\bibnamefont{Kotar}},
  \bibinfo{author}{\bibfnamefont{L.}~\bibnamefont{Debono}},
  \bibinfo{author}{\bibfnamefont{N.}~\bibnamefont{Bruot}},
  \bibinfo{author}{\bibfnamefont{S.}~\bibnamefont{Box}},
  \bibinfo{author}{\bibfnamefont{D.}~\bibnamefont{Phillips}},
  \bibinfo{author}{\bibfnamefont{S.}~\bibnamefont{Simpson}},
  \bibinfo{author}{\bibfnamefont{S.}~\bibnamefont{Hanna}}, \bibnamefont{and}
  \bibinfo{author}{\bibfnamefont{P.}~\bibnamefont{Cicuta}},
  \bibinfo{journal}{Phys.~Rev.~Lett.} \textbf{\bibinfo{volume}{111}},
  \bibinfo{pages}{228103} (\bibinfo{year}{2013}).

\bibitem[{\citenamefont{Arzola et~al.}(2014)\citenamefont{Arzola, Jákl,
  Chvátal, and Zemánek}}]{Arzola.Zemanek2014}
\bibinfo{author}{\bibfnamefont{A.~V.} \bibnamefont{Arzola}},
  \bibinfo{author}{\bibfnamefont{P.}~\bibnamefont{Jákl}},
  \bibinfo{author}{\bibfnamefont{L.}~\bibnamefont{Chvátal}}, \bibnamefont{and}
  \bibinfo{author}{\bibfnamefont{P.}~\bibnamefont{Zemánek}},
  \bibinfo{journal}{Opt. Express} \textbf{\bibinfo{volume}{22}},
  \bibinfo{pages}{16207} (\bibinfo{year}{2014}).

\bibitem[{\citenamefont{Mihiretie et~al.}(2012)\citenamefont{Mihiretie, Snabre,
  Loudet, and Pouligny}}]{Mihiretie.Pouligny2012}
\bibinfo{author}{\bibfnamefont{B.~M.} \bibnamefont{Mihiretie}},
  \bibinfo{author}{\bibfnamefont{P.}~\bibnamefont{Snabre}},
  \bibinfo{author}{\bibfnamefont{J.~C.} \bibnamefont{Loudet}},
  \bibnamefont{and} \bibinfo{author}{\bibfnamefont{B.}~\bibnamefont{Pouligny}},
  \bibinfo{journal}{Europhys.~Lett.} \textbf{\bibinfo{volume}{100}},
  \bibinfo{pages}{48005} (\bibinfo{year}{2012}).

\bibitem[{\citenamefont{Mihiretie et~al.}(2013)\citenamefont{Mihiretie, Loudet,
  and Pouligny}}]{Mihiretie.Pouligny2013}
\bibinfo{author}{\bibfnamefont{B.}~\bibnamefont{Mihiretie}},
  \bibinfo{author}{\bibfnamefont{J.-C.} \bibnamefont{Loudet}},
  \bibnamefont{and} \bibinfo{author}{\bibfnamefont{B.}~\bibnamefont{Pouligny}},
  \bibinfo{journal}{J. Quant.\ Spectrosc.\ Radiat.\ Transf.}
  \textbf{\bibinfo{volume}{126}}, \bibinfo{pages}{61 } (\bibinfo{year}{2013}).

\bibitem[{\citenamefont{Kavčič et~al.}(2012)\citenamefont{Kavčič, Babić,
  Osterman, Podobnik, and Poberaj}}]{Kavcic.Poberaj2012}
\bibinfo{author}{\bibfnamefont{B.}~\bibnamefont{Kavčič}},
  \bibinfo{author}{\bibfnamefont{D.}~\bibnamefont{Babić}},
  \bibinfo{author}{\bibfnamefont{N.}~\bibnamefont{Osterman}},
  \bibinfo{author}{\bibfnamefont{B.}~\bibnamefont{Podobnik}}, \bibnamefont{and}
  \bibinfo{author}{\bibfnamefont{I.}~\bibnamefont{Poberaj}},
  \bibinfo{journal}{Microsyst.\ Technol.} \textbf{\bibinfo{volume}{18}},
  \bibinfo{pages}{191} (\bibinfo{year}{2012}).

\bibitem[{\citenamefont{Kotar et~al.}(2006)\citenamefont{Kotar, Vilfan,
  Osterman, Babi\v{c}, \v{C}opi\v{c}, and Poberaj}}]{Kotar.Poberaj2006}
\bibinfo{author}{\bibfnamefont{J.}~\bibnamefont{Kotar}},
  \bibinfo{author}{\bibfnamefont{M.}~\bibnamefont{Vilfan}},
  \bibinfo{author}{\bibfnamefont{N.}~\bibnamefont{Osterman}},
  \bibinfo{author}{\bibfnamefont{D.}~\bibnamefont{Babi\v{c}}},
  \bibinfo{author}{\bibfnamefont{M.}~\bibnamefont{\v{C}opi\v{c}}},
  \bibnamefont{and} \bibinfo{author}{\bibfnamefont{I.}~\bibnamefont{Poberaj}},
  \bibinfo{journal}{Phys.~Rev.~Lett.} \textbf{\bibinfo{volume}{96}},
  \bibinfo{pages}{207801} (\bibinfo{year}{2006}).

\bibitem[{\citenamefont{Osterman}(2010)}]{Osterman2010}
\bibinfo{author}{\bibfnamefont{N.}~\bibnamefont{Osterman}},
  \bibinfo{journal}{Comput.\ Phys.\ Commun.} \textbf{\bibinfo{volume}{181}},
  \bibinfo{pages}{1911} (\bibinfo{year}{2010}).

\bibitem[{\citenamefont{Happel and Brenner}(1983)}]{Happel.Brenner}
\bibinfo{author}{\bibfnamefont{J.}~\bibnamefont{Happel}} \bibnamefont{and}
  \bibinfo{author}{\bibfnamefont{H.}~\bibnamefont{Brenner}},
  \emph{\bibinfo{title}{Low Reynolds Number Hydrodynamics}}
  (\bibinfo{publisher}{Kluwer, Dodrecht}, \bibinfo{year}{1983}).

\bibitem[{\citenamefont{Blake}(1971)}]{Blake.1971}
\bibinfo{author}{\bibfnamefont{J.~R.} \bibnamefont{Blake}},
  \bibinfo{journal}{Proc.~Camb.~Phil.~Soc.} \textbf{\bibinfo{volume}{70}},
  \bibinfo{pages}{303} (\bibinfo{year}{1971}).

\bibitem[{\citenamefont{Pozrikidis}(2002)}]{Pozrikidis2002}
\bibinfo{author}{\bibfnamefont{C.}~\bibnamefont{Pozrikidis}},
  \emph{\bibinfo{title}{A practical guide to boundary element methods with the
  software library {BEMLIB}}} (\bibinfo{publisher}{CRC Press},
  \bibinfo{year}{2002}).

\bibitem[{\citenamefont{Lenz and Ryskin}(2006)}]{Lenz.Ryskin2006}
\bibinfo{author}{\bibfnamefont{P.}~\bibnamefont{Lenz}} \bibnamefont{and}
  \bibinfo{author}{\bibfnamefont{A.}~\bibnamefont{Ryskin}},
  \bibinfo{journal}{Phys.~Biol.} \textbf{\bibinfo{volume}{3}},
  \bibinfo{pages}{285} (\bibinfo{year}{2006}).

\bibitem[{\citenamefont{Wollin and Stark}(2011)}]{Wollin.Stark2011}
\bibinfo{author}{\bibfnamefont{C.}~\bibnamefont{Wollin}} \bibnamefont{and}
  \bibinfo{author}{\bibfnamefont{H.}~\bibnamefont{Stark}},
  \bibinfo{journal}{Eur.\ Phys.\ J. E Soft Matter}
  \textbf{\bibinfo{volume}{34}}, \bibinfo{pages}{1} (\bibinfo{year}{2011}).

\end{thebibliography}
\end{document}